\documentclass[superscriptaddress,showkeys,showpacs,preprintnumbers,floatfix,twocolumn]{revtex4-1}

\RequirePackage{lineno}
\usepackage{epsfig}
\usepackage{graphicx}
\usepackage{pslatex}
\usepackage{color}
\usepackage{amssymb}
\usepackage{multirow}
\usepackage{xspace}

\newcommand{\Ups}{\ensuremath{\Upsilon}\xspace}
\newcommand{\UpsAny}{\ensuremath{{\Upsilon(n\text{S})}}\xspace}
\newcommand{\UpsOne}{\ensuremath{{\Upsilon(\text{1S})}}\xspace}
\newcommand{\UpsTwo}{\ensuremath{{\Upsilon(\text{2S})}}\xspace}
\newcommand{\UpsThree}{\ensuremath{{\Upsilon(\text{3S})}}\xspace}
\newcommand{\UpsExc}{\ensuremath{{\Upsilon(\text{2S+3S})}}\xspace}
\newcommand{\UpsAll}{\ensuremath{{\Upsilon(\text{1S+2S+3S})}}\xspace}
\newcommand{\Jpsi}{\ensuremath{J/\psi}\xspace}
\newcommand{\pT}{\ensuremath{p_\mathrm{T}}\xspace}

\newcommand{\Raa}{\ensuremath{R_\mathrm{AA}}\xspace}
\newcommand{\Npart}{\ensuremath{N_\mathrm{part}}\xspace}
\newcommand{\Ncoll}{\ensuremath{N_\mathrm{coll}}\xspace}
\newcommand{\Etwr}{\ensuremath{E_\mathrm{tower}}\xspace}
\newcommand{\Eclus}{\ensuremath{E_\mathrm{cluster}}\xspace}
\newcommand{\nsige}{\ensuremath{{n\sigma_{e}}}\xspace}

\newcommand{\Bee}{\ensuremath{B_{ee}}\xspace}
\newcommand{\ee}{\ensuremath{e^{+}e^{-}}\xspace}
\newcommand{\mub}{\ensuremath{\mu\mathrm{b}}\xspace}

\newcommand{\bb}{\ensuremath{b\bar{b}}\xspace}

\begin{document}

\title{\Ups production in U+U collisions at $\sqrt{s_{NN}}=193$ GeV with the STAR experiment}
%
\author{L.~Adamczyk}\affiliation{AGH University of Science and Technology, FPACS, Cracow 30-059, Poland}
\author{J.~K.~Adkins}\affiliation{University of Kentucky, Lexington, Kentucky, 40506-0055}
\author{G.~Agakishiev}\affiliation{Joint Institute for Nuclear Research, Dubna, 141 980, Russia}
\author{M.~M.~Aggarwal}\affiliation{Panjab University, Chandigarh 160014, India}
\author{Z.~Ahammed}\affiliation{Variable Energy Cyclotron Centre, Kolkata 700064, India}
\author{I.~Alekseev}\affiliation{Alikhanov Institute for Theoretical and Experimental Physics, Moscow 117218, Russia}\affiliation{National Research Nuclear University MEPhI, Moscow 115409, Russia}
\author{D.~M.~Anderson}\affiliation{Texas A\&M University, College Station, Texas 77843}
\author{R.~Aoyama}\affiliation{Brookhaven National Laboratory, Upton, New York 11973}
\author{A.~Aparin}\affiliation{Joint Institute for Nuclear Research, Dubna, 141 980, Russia}
\author{D.~Arkhipkin}\affiliation{Brookhaven National Laboratory, Upton, New York 11973}
% ...

\collaboration{STAR Collaboration}\noaffiliation

\begin{abstract}
We present a measurement of the inclusive production of \Ups mesons in
U+U collisions at $\sqrt{s_{NN}}=193$ GeV at mid-rapidity ($|y|<1$). 
Previous studies in central Au+Au collisions at
$\sqrt{s_{NN}}=200$ GeV show a suppression of \UpsAll
production relative to expectations from the \Ups yield in {\it{p+p}} collisions scaled by the number
of binary nucleon-nucleon collisions (\Ncoll), with an indication that the \UpsOne state is also suppressed.
The present measurement extends the number of participant nucleons
in the collision (\Npart) by 20\% compared to Au+Au collisions, and allows us to study a system with higher energy density. 
We observe a suppression in both the \UpsAll{} and \UpsOne{} yields in central U+U data, which consolidates and extends the previously observed suppression trend in Au+Au collisions.


\end{abstract}

\maketitle

\section{Introduction}
Quarkonium production in high energy heavy-ion collisions is
expected to be sensitive to the energy density and temperature of the medium created in these collisions.
Dissociation of different quarkonium states due to color
screening is predicted to depend on their binding
energies~\cite{Digal:2001iu,Wong:2004zr,Cabrera:2006nt}.
Measuring the yields of different quarkonium states 
therefore may serve as a model-dependent measure of the temperature in the medium~\cite{Mocsy:2007jz}. Although charmonium suppression was
anticipated as a key signature of the formation of a quark-gluon
plasma (QGP)~\cite{Matsui:1986dk}, the suppression of $J/\psi$ mesons has been found to be relatively independent
of beam energy from Super Proton Synchrotron (SPS) to Relativistic
Heavy Ion Collider (RHIC) energies~\cite{Adare:2006ns}. This
phenomenon can be attributed to 
$J/\psi$ regeneration by the recombination of uncorrelated 
$c$-$\bar{c}$ pairs in the deconfined medium~\cite{Grandchamp:2004tn} that counterbalances the dissociation process.
In addition, cold nuclear matter (CNM) effects, dissociation in the
hadronic phase, and feed-down contributions from excited charmonium
states and $B$ hadrons can alter the suppression pattern from what would be expected from Debye screening. Contrary to the more abundantly produced charm quarks, bottom pair
recombination and co-mover absorption effects are predicted to be negligible at RHIC energies~\cite{Rapp:2008tf}. Bottomonium states in heavy-ion
collisions therefore can serve as a cleaner probe of the medium,
although initial state effects may still play an important
role~\cite{Grandchamp:2005yw,Adamczyk:2013poh,afraw,Vogt:2012fba,Arleo:2012rs}.
Feed-down from $\chi_b$ mesons, the yield of which is largely unknown at RHIC energies, may also give a non-negligible contribution to the bottomonium yields.

Monte Carlo Glauber simulations show that collisions of large,
deformed uranium nuclei reach on average a higher number of
participant nucleons (\Npart)
and higher number of binary nucleon-nucleon collisions (\Ncoll) than gold-gold collisions of
the same centrality class. It was estimated that central U+U
collisions at $\sqrt{s_{NN}}$=193 GeV have an
approximately 20\% higher energy density, thus higher temperature, than that in central Au+Au
collisions at $\sqrt{s_{NN}}$=200 GeV~\cite{Kikola:2011zz,Mitchell:2013mza}. Lattice quantum-chromodynamics (QCD) calculations at finite temperature suggest that the color screening radius decreases
with increasing temperature as $r_D(T) \sim 1/T$, which implies that a
given quarkonium state cannot form above a certain
temperature threshold~\cite{Petrov:2007zza}. Free-energy-based
spectral function calculations predict that the excited \UpsExc states
cannot exist above $1.2 T_c$ and that the ground state \UpsOne{} cannot exist above approximately $2 T_c$, where $T_c$ is the critical temperature of the phase transition~\cite{Mocsy:2007jz}. Around the onset of deconfinement, one may see a sudden drop in the production of a given \Ups state when the threshold temperature of that state (or of higher mass states that decay into it) is reached. 
According to Ref.~\cite{Kikola:2011zz}, in the 5\% most central U+U collisions at $\sqrt{s_{NN}}=193$ GeV, $T/T_c$ is between 2 and 2.7, depending on the \Ups{} formation time chosen in calculations. For a given formation time, the value of $T/T_c$ is approximately 5\% higher than in the 5\% most central Au+Au collisions at $\sqrt{s_{NN}}=200$ GeV.
In such a scenario the temperature present in central U+U
collisions is high enough that even the \UpsOne state might dissociate.
However, the finite size, lifetime and inhomogeneity of the plasma may complicate this picture and smear the turn-on of the melting of particular
quarkonium states over a wide range of \Npart.
The suppression of bottomonium states in U+U collisions, together with existing measurements in other
collision systems as well as measurements of CNM effects, may provide the means to explore the
turn-on characteristics of suppression and test the sequential melting hypothesis.

\section{Experiment and analysis}

This analysis uses data recorded in 2012 by the STAR experiment at
RHIC in U+U collisions at
$\sqrt{s_{NN}}=193$ GeV. 

\begin{figure}[!t]
  \includegraphics[trim=0 0 17mm 0,clip,width=\linewidth]{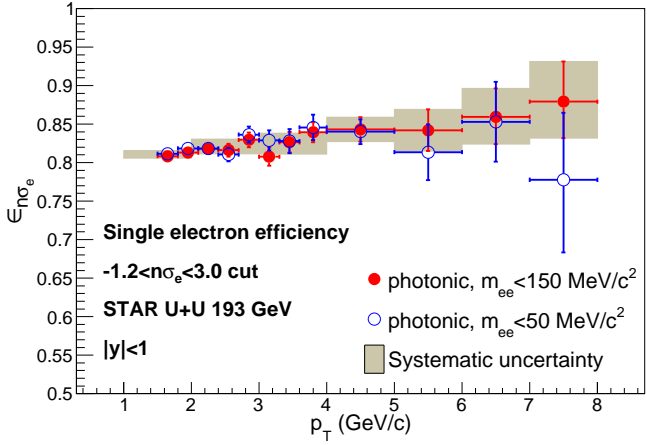}
  \caption{\label{fig:nSigE} {\it (Color online)} 
Single electron efficiency of the $dE/dx$ cut versus transverse
momentum, as determined by fits to \nsige distributions of photonic electrons. The fit errors using the sample with the $m_{ee}<150$ MeV/$c^2$ photonic electron cut in 1 GeV/$c$ wide bins are used as systematic
uncertainties. The results using the $m_{ee}<50$ MeV/$c^2$ photonic electron cut are consistent with the former one.}
\end{figure}

We reconstruct the \Ups{} states via their dielectron decay channels, $\Ups\rightarrow\ee$, based on the method decribed in Ref.~\cite{Adamczyk:2013poh}.
As a trigger we require at least one tower from the Barrel Electromagnetic Calorimeter (BEMC)~\cite{Beddo:2002zx} within the pseudorapidity range $|\eta|<1$, containing a signal corresponding to an energy deposit that is higher than approximately 4.2 GeV. A total of 17.2 million BEMC-triggered events are analyzed, corresponding to an integrated luminosity of 263.4 $\mu$b$^{-1}$.
The electron (or positron) candidate that caused the trigger signal is paired with other electron candidates within the same event.
Tracks are reconstructed in the Time Projection Chamber
(TPC)~\cite{Ackermann:2002ad}. Electrons with a momentum $p>1.5$
GeV/$c$ are selected based on their specific energy loss ($dE/dx$) in the TPC. Candidates are required to lie within an asymmetric window of $-1.2<\nsige<3$, 
where \nsige is the deviation of the measured $dE/dx$ with respect to the nominal $dE/dx$ value for electrons at a given momentum, calculated using the Bichsel parametrization~\cite{Bichsel:2006cs}, normalized with the TPC resolution.
Figure~\ref{fig:nSigE} shows 
the efficiency of the \nsige cut ($\epsilon_\nsige$) for single electrons versus transverse momentum (\pT), determined using a high purity electron sample obtained from gamma conversions.
Since most of these so-called photonic electron pairs are contained in the very low invariant mass ($m_{ee}$) regime, we select $\ee$ pairs with $m_{ee}<150$ MeV/$c^2$ ($m_{ee}<50$ MeV/$c^2$ in systematics checks) in a similar manner to the analysis described in Ref.~\cite{Agakishiev:2011mr}.

\begin{figure}[!t]
  \includegraphics[trim=0 0 17mm 0,clip,width=\linewidth]{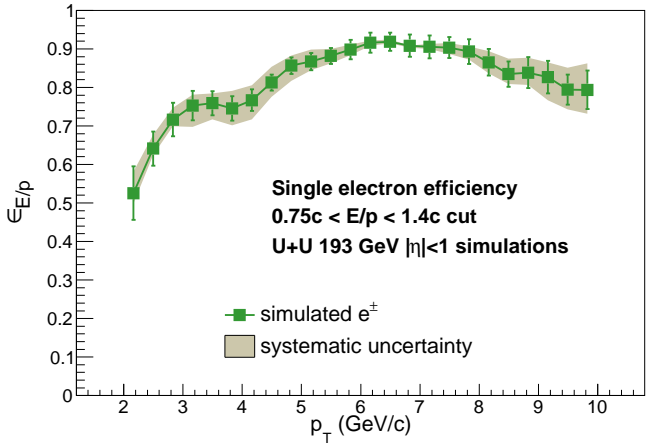}
  \caption{\label{fig:EoP} {\it (Color online)} 
Single electron efficiency of the $\Eclus/p$ cut versus
transverse momentum. The efficiency corrections are obtained from embedded
simulations. The difference between the default result from simulations and that extracted using a pure electron sample from data is taken as
the systematic uncertainty.}
\end{figure}

To further enhance the purity of the electron sample we use the particle discrimination power of the BEMC.
Electromagnetic showers tend to be more compact than hadron showers, and deposit their energy in fewer towers. The total energy
deposit of an electron candidate (\Eclus) is determined by finding a {\it seed} 
tower with a high energy deposit (\Etwr), and forming a {\it cluster} by joining the two highest-energy neighbours to this seed. 
An $R=\sqrt{\Delta\varphi^2+\Delta\eta^2}<0.04$ matching cut
is applied on the distance of the 
seed tower position in the BEMC and the TPC track projected to the BEMC plane, expressed in azimuthal angle and pseudorapidity units. We reconstruct the quantity $\Eclus/p$ for each electron candidate, where $p$ is the momentum of the electron candidate measured in the TPC. Electrons travelling close to the speed of light are expected to follow an $\Eclus/p$ distribution centered at $c$, smeared by the TPC and BEMC detector resolutions. Therefore a $0.75 c
<\Eclus/p<1.4 c$ cut is applied to reject hadron
background. The efficiency of this cut for single electrons ($\epsilon_{E/p}$), obtained from detector simulation studies, is shown in Fig.~\ref{fig:EoP}.
Since the trigger is already biased towards more compact clusters, an \Ups candidate requires that the daughter electron candidate that fired the trigger fulfills a strict condition of $\Etwr/\Eclus>0.7$, while the daughter paired to it is required to fulfill a looser $\Etwr/\Eclus>0.5$ cut.

%
\begin{figure}[!t]
  \includegraphics[trim=0 0 17mm 0,clip,width=\linewidth]{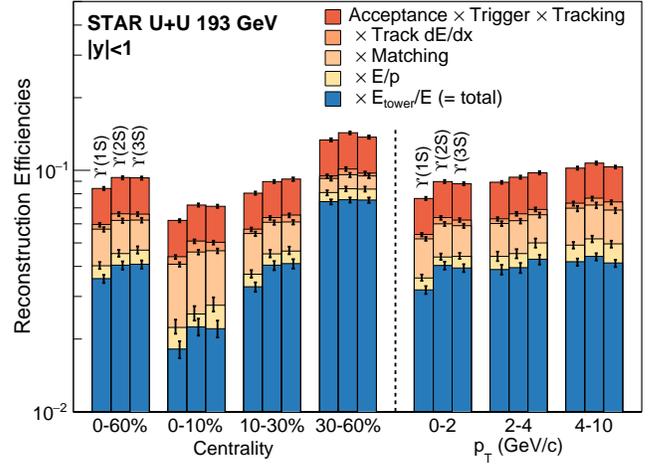}
  \caption{\label{fig:effs} {\it (Color online)} 
Reconstruction efficiencies for \UpsOne{}, \UpsTwo{} and \UpsThree{}, as
determined from embedded simulations and identified electron samples. Cuts for {\it i)} acceptance, triggering and tracking, {\it ii)} specific energy loss, {\it iii)} track--cluster matching, {\it iv)} $\Eclus/p$ and {\it v)} cluster compactness (\Etwr/\Eclus) are applied consecutively to build up the total reconstruction efficiency. The efficiencies corresponding to each cut are shown stacked in a top-to-bottom order. Black ticks at the end of each bar represent the total uncertainties on the given efficiency.
The \pT{}-binned values correspond to 0-60\% centrality.
}
\end{figure}

The acceptance, as well as the tracking, the triggering and the BEMC cut
efficiency correction factors are determined using simulations,
where the $\UpsAny\rightarrow\ee$ processes ($n$=1,2,3) are embedded
into U+U collision events, and then reconstructed in the same way as real data. The efficiency of the $dE/dx$ cut is determined by using the single electron efficiency from photonic electrons, as shown in Fig.~\ref{fig:nSigE}. The BEMC-related reconstruction efficiencies are also verified with a sample of electrons identified in the TPC.
Figure~\ref{fig:effs} shows the
reconstruction efficiencies for \UpsOne, \UpsTwo and \UpsThree
states separately, for 0--60\% centrality, as well as for centrality
bins 0--10\%, 10--30\%, 30--60\%, and transverse momentum bins of $\pT<2$ GeV/$c$, $2<\pT<4$ GeV/$c$ and $4<\pT<10$ GeV/$c$. 
%

\begin{figure}
  \includegraphics[trim=0 0 17mm 0,clip,width=\linewidth]{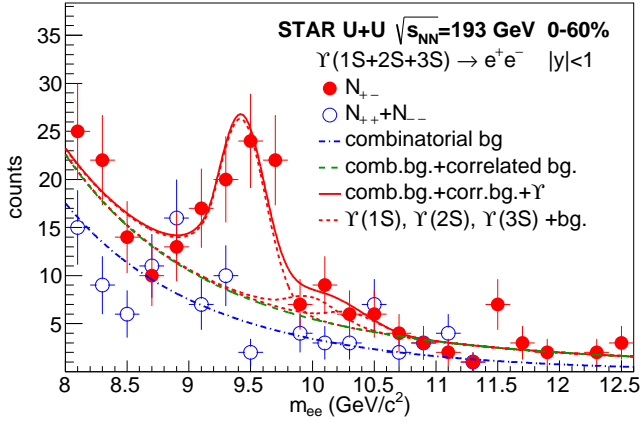}
  \caption{\label{fig:invmass} {\it (Color online)}     
Reconstructed invariant mass distribution of \Ups{} candidates (unlike-sign pairs, denoted as solid circles) and like-sign combinatorial background (open circles)
in U+U collisions at $\sqrt{s_{NN}}=193$ GeV for 0-60\% centrality at mid-rapidity ($|y|<1$).
Fits to the combinatorial background, \bb and Drell-Yan contributions
and to the \Ups peaks are plotted as dash-dotted, dashed and
solid lines respectively. The fitted contributions of the individual \UpsOne, \UpsTwo and \UpsThree states are shown as dotted
lines.}
\end{figure}

\begin{figure}
  \includegraphics[trim=2mm 0 17mm 0,clip,width=\linewidth]{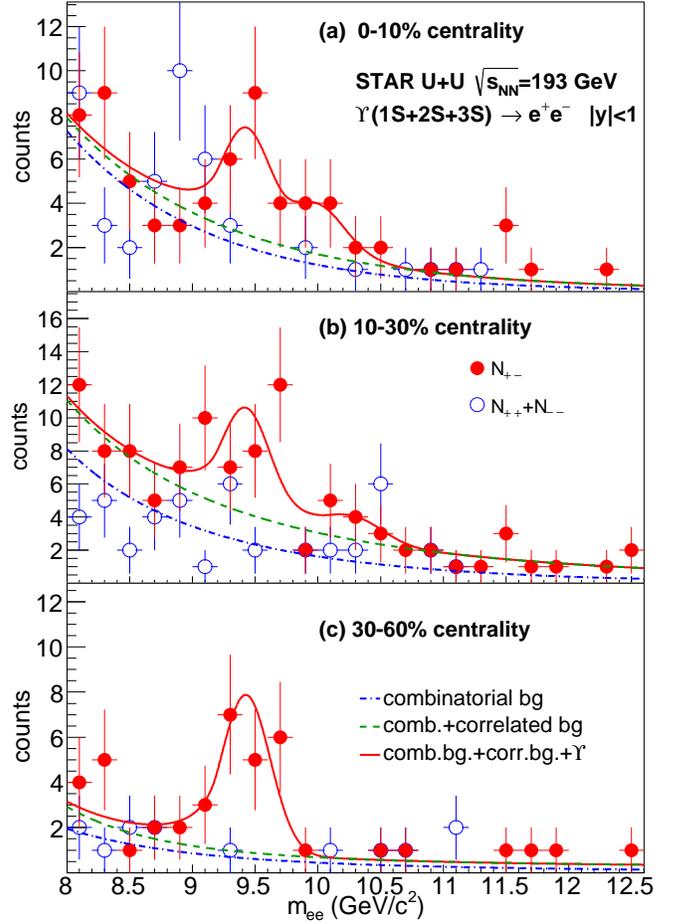}%
  \caption{\label{fig:invmasspanels} {\it (Color online)}     
Reconstructed invariant mass distribution of \Ups{} candidates (solid circles) and like-sign combinatorial background (open circles)
in U+U collisions at $\sqrt{s_{NN}}=193$ GeV for \pT-integrated 0-10\% (a)
10-30\% (b) 30-60\% (c) centralities at mid-rapidity
($|y|<1$). Fits to the combinatorial background, \bb and Drell-Yan contributions and the peak fits are plotted as
dash-dotted, dashed and
solid lines respectively.}
\end{figure}

The invariant mass spectrum of the \Ups candidates is reconstructed within the rapidity window $|y|<1$ using dielectron momenta measured in the TPC. Figure~\ref{fig:invmass} shows the $m_{ee}$ distribution of unlike-sign pairs as solid circles, along with the sum of the positive and negative like-sign distributions as open circles. The data are divided into three centrality bins, shown in Fig.~\ref{fig:invmasspanels}, 
and three \pT bins.
%
%
The measured signal from each of the $\Upsilon(nS)\rightarrow\ee$
processes ($n=1,2,3$) is parametrized with a Crystal Ball
function~\cite{Gaiser:1982yw}, with parameters obtained from fits to the \UpsAny mass peaks from simulations. Such a shape was justified by
preceding studies~\cite{Abelev:2010am} and accounts for the effects of Bremsstrahlung and the momentum resolution of the TPC.
The combinatorial background is modelled with a double exponential
function. In addition, there is a sizeable correlated background from \bb{} decays and Drell-Yan processes. Based on previous studies~\cite{Adamczyk:2013poh,Abelev:2010am} we use a ratio of two power law functions that were found to adequately describe these contributions. In order to determine the $\Upsilon$ yield, a simultaneous log-likelihood fit is performed on the like-sign and the unlike-sign data. The unlike-sign data are fitted with a function that includes the combinatorial and correlated background shapes plus the three \Ups mass peaks, while the like-sign data is fitted with the combinatorial background shape only.
The parameters of the mass peaks and those of the correlated background are fixed in the fit according to the simulations and previous studies~\cite{Adamczyk:2013poh,Abelev:2010am}, respectively, except for normalization parameters.
The contribution of each \UpsAny state to the total \UpsAll{} yield is determined based on the integral of the individual Crystal Ball functions that are fit to the measured peaks. The uncertainties quoted as statistical are the uncertainties from the fit.

\section{Systematic uncertainties}

We consider several sources of systematic uncertainties in the present
study. Geometrical acceptance is affected by \Ups{} polarization as well as by
noisy towers that are not used in the reconstruction. The sytematics
stemming from these factors, estimated in Ref.~\cite{Adamczyk:2013poh},
are taken as fully correlated between collision systems. 
The geometrical acceptance correction factor is dependent on the \pT and rapidity distributions of the \Ups mesons.
We assume a Boltzmann-like \pT{}-distribution, $\frac{dN}{d\pT}
  \propto \frac{\pT}{\exp({\pT}/{p_0})+1}$, in our embedded
  simulations. We obtain its slope parameter of $p_0=1.11$ GeV/$c$ from a 
parametrized interpolation of {\it{p+p}} data from ISR, CDF and 
measurements~\cite{Acosta:2001gv,Kourkoumelis:1980hg,Khachatryan:2010zg}, similar to Ref.~\cite{Adamczyk:2013poh}.
Although this value matches the fit to the \pT spectrum of the current analysis, detailed in Sec.~\ref{sec:results}, there is a slight difference between the two within the statistical error range. The uncertainty from the slope is determined by adjusting the slope to match the fitted value, $p_0=1.37$ GeV/$c$.
The rapidity distribution is determined using
PYTHIA~\cite{Sjostrand:1993yb} version 8.1 to follow an approximately
Gaussian shape with $\sigma=1.15$. We vary the width between 1.0 and 1.16 to cover the range of the uncertainties of the Gaussian fit, as well as estimations of earlier studies~\cite{Adamczyk:2013poh}.

The uncertainty of the TPC track reconstruction efficiency caused
by the variation in operational conditions was studied in Refs.~\cite{Adamczyk:2013poh,Adler:2001yq}.
The errors of the Gaussian fits to the \nsige 
distribution of photonic electrons are taken as the uncertainties on the electron identification using TPC ($dE/dx$).
Changing the photonic electron selection from the default $m_\mathrm{ee}<150$ MeV/$c^2$ to $m_{ee}<50$ MeV/$c^2$, or using TPC-identified
electrons instead of photonic ones yield a result that is consistent
with the default choice within systematic uncertainties. Figure~\ref{fig:nSigE} shows the systematic uncertainty corresponding to the $dE/dx$ single electron efficiency as a band around the data points.
The uncertainty stemming from the trigger turn-on characteristics, from the criteria of electron selection with the BEMC (matching, $\Eclus/p$, as well as
the cluster compactness $\Etwr/\Eclus$) are determined from the comparison of efficiencies calculated from embedded simulations and from electron samples obtained from data using TPC ($dE/dx$) identification and
reconstructed photonic conversion electrons. 
The dominant source of systematic uncertainty among those listed above is the uncertainty of the $\Eclus/p$ cut efficiency.
In Fig.~\ref{fig:EoP} we indicate the systematic uncertainty corresponding to the single electron $\Eclus/p$ efficiency with a band around the data points.

Another major source of uncertainty arises from the assumptions of the signal and background shapes made in extracting the signal yield. The extraction method was systematically modified to
estimate the uncertainties from momentum resolution and
calibration, functional shapes of the correlated and combinatorial backgrounds as well as the signal, and
those from the fit range in the following ways: {\it i)} An additional 50 MeV/$c^2$ smearing was
added to the peaks to model a worst-case scenario in the momentum
resolution~\cite{Abelev:2010am}; {\it ii)} The double exponential fit function used for the combinatorial background was replaced with a single exponential function; {\it iii)} 
Instead of modelling the correlated background with a ratio of two power law functions, we used a single power law function to commonly represent the Drell-Yan and \bb contributions, and we also tested the sum of these two functions to represent the Drell-Yan and \bb contributions individually in the fitting;
{\it iv)} Finally, we moved the lower and upper limits of the simultaneous fit range in several steps from 6.6 to 8.0 GeV/$c^2$ and from 15.4 to 12.4 GeV/$c^2$
respectively. The \Ups yields were determined in each case, and the
maximum deviation from the default case in positive or negative
direction was taken as the signal extraction uncertainty.

We construct the nuclear modification factor, \Raa, to quantify the medium effects on the 
production of the \Ups states. The \Raa is computed by comparing the corrected
number of \Ups mesons measured in A+A collisions to the yield in {\it{p+p}}
collisions scaled by the average number of binary nucleon-nucleon collisions, as
$\Raa^\Upsilon=\frac{\sigma^{inel}_{pp}}{\sigma^{inel}_{AA}}\frac{1}{\langle\Ncoll\rangle}\frac{\Bee\times(d\sigma^{AA}_\Ups
  / dy)}{\Bee\times(d\sigma^{pp}_\Ups
  / dy)}$
, where $\sigma^{inel}_{AA(pp)}$ is the total
inelastic cross-section of the U+U ({\it{p+p}}) collisions,
$d\sigma^{AA(pp)}_\Ups /dy$ denotes the \Ups
production cross-section in U+U ({\it{p+p}}) collisions, and $\Bee$ is the
branching ratio of the $\Upsilon\rightarrow\ee$ process.
Our reference was measured in {\it{p+p}} collisions at $\sqrt{s}=200$
GeV~\cite{Abelev:2010am}, and has to be scaled to $\sqrt{s}=193$ GeV.
Calculations for the {\it{p+p}} inelastic cross-section~\cite{Schuler:1993wr} yield a 0.5\% smaller value at $\sqrt{s}=193$ GeV than at
$\sqrt{s}=200$ GeV. The \Ups production cross-section, however,
shows a stronger dependence on the collision energy. Both the NLO color-evaporation model calculations, which describe the world {\it{p+p}} data~\cite{Bedjidian:2004gd}, and a linear interpolation of the same data points
within the RHIC-LHC energy regime yield an approximately 4.6\% decrease in the cross
section when $\sqrt{s}$ is changed from 200 to 193 GeV.
The uncertainties do not exceed 0.5\% (absolute) 
in any of these corrections, and are thus neglected.
The values used to compute \Raa are $\left. \Bee\times(d\sigma^{pp}_\Ups/dy)\right|_{|y|<1}=60.64$ pb, $\sigma^{inel}_{pp} = 42.5$ mb and $\sigma^{inel}_{UU} = 8.14$ b.
The \Npart and \Ncoll values used in this analysis, computed using the Monte Carlo Glauber model~\cite{Alver:2008aq} following the method of Ref.~\cite{Masui:2009qk}, are listed in Table~\ref{tab:glauber}.
\begin{table}
\center
\begin{tabular}{| c | c | c |}
\hline
{centrality} & \Npart & \Ncoll \\
\hline
\hline
0-60\%  & 188.3$\pm$5.5 & 459$\pm$10 \\
0-10\%  & 385.1$\pm$9   & 1146$\pm$49 \\
10-30\% & 236.2$\pm$14  & 574$\pm$41 \\
30-60\% & 91.0$\pm$32  & 154$\pm$37 \\
\hline
\end{tabular}
\caption{\label{tab:glauber}The \Ncoll and \Npart values corresponding to different centrality ranges, obtained using the Monte Carlo Glauber model.}
\end{table}

The systematic uncertainties for U+U collisions at 0-60\% centrality are summarized in Table~\ref{tab:syst}.
The total relative systematic uncertainty on $\Raa^\Ups$,
calculated as a quadratic sum of the uncertainties listed in the table excluding common normalization uncertainties from the {\it p+p} reference measurements, ranges from 15\% to 27\%
dependent on centrality and \pT{}.
\begin{table}
\center
\begin{tabular}{| c | c | c |}
\hline
\multicolumn{2}{|c|}{Source of systematic uncertainty} & value (\%) \\
\hline
\hline
\multicolumn{2}{|c|}{Number of binary collisions (\Raa-only)} & 2.2 \\
\multicolumn{2}{|c|}{Geometrical acceptance  (yield-only)} & $^{+1.7}_{-3.0}$\\
\multicolumn{2}{|c|}{\pT and $y$ distributions} & 2.1 \\
\multicolumn{2}{|c|}{Trigger efficiency} & $^{+1.1}_{-3.6}$\\
\multicolumn{2}{|c|}{Tracking efficiency} & 11.8\\
\multicolumn{2}{|c|}{TPC $dE/dx$} & $^{+4.0}_{-6.4}$\\
\multicolumn{2}{|c|}{TPC--BEMC matching} & 5.4 \\
\multicolumn{2}{|c|}{BEMC $\Eclus/p$} & $^{+8.8}_{-13.2}$\\
\multicolumn{2}{|c|}{BEMC $\Etwr/\Eclus$} & 2.0 \\
\hline
\multirow{3}{*}{Signal extraction} & \UpsAll & $^{+8.4}_{-7.0}$ \\
  & \UpsOne & $^{+11.9}_{-5.7}$ \\
  & \UpsExc &  $^{+5.3}_{-19.7}$ \\
\hline
\end{tabular}
\caption{\label{tab:syst}Major systematic uncertainties excluding
common normalization uncertainties from the {\it{p+p}} reference, for 0-60\% centrality data.}
\end{table}

\section{Results}
\label{sec:results}

The production cross-sections are summarized in Table~\ref{tab:result} for the sum of all three \Ups{} states, the separated \UpsOne state, and for the excited \UpsExc states together.
\begin{table}
\center
\begin{tabular}{| c | c | c | c |}
\hline
states & centrality & $\Bee \times (d\sigma^\Upsilon_{AA} /dy)$ ($\mu$b) & $\Raa^\Upsilon$ \\
\hline

\hline
\multirow{4}{*}{\UpsAll}
 & 0--60\% & $4.27 \pm 0.90^{+0.90}_{-0.82}$ & $0.82 \pm 0.17^{+0.14}_{-0.11}$ \\
 & 0--10\% & $6.64 \pm 4.22^{+1.95}_{-1.66}$ & $0.51 \pm 0.32^{+0.13}_{-0.11}$ \\
 & 10--30\% & $3.67 \pm 1.62^{+1.04}_{-0.78}$ & $0.56 \pm 0.25^{+0.14}_{-0.10}$ \\
 & 30--60\% & $3.42 \pm 1.04^{+0.57}_{-0.97}$ & $1.96 \pm 0.59^{+0.51}_{-0.68}$ \\
\hline
\multirow{4}{*}{\UpsOne}
 & 0--60\% & $3.55 \pm 0.77^{+0.80}_{-0.66}$ & $0.96 \pm 0.21^{+0.18}_{-0.13}$ \\
 & 0--10\% & $4.52 \pm 2.08^{+1.31}_{-1.13}$ & $0.49 \pm 0.23^{+0.12}_{-0.10}$ \\
 & 10--30\% & $2.91 \pm 1.10^{+0.85}_{-0.61}$ & $0.63 \pm 0.24^{+0.17}_{-0.11}$ \\
 & 30--60\% & $3.42 \pm 0.95^{+0.57}_{-0.97}$ & $2.76 \pm 0.76^{+0.71}_{-0.95}$ \\
\hline
\multirow{3}{*}{\UpsExc}
 & 0--60\% & $0.72 \pm 0.49^{+0.15}_{-0.19}$ & $0.48 \pm 0.32^{+0.07}_{-0.11}$ \\
 & 0--10\% & $2.11 \pm 3.33^{+0.64}_{-0.54}$ & $0.56 \pm 0.89^{+0.15}_{-0.12}$ \\
 & 10--30\% & $0.76 \pm 1.03^{+0.29}_{-0.16}$ & $0.41 \pm 0.55^{+0.15}_{-0.07}$ \\
\hline

\end{tabular}
\caption{\label{tab:result}Cross-sections multiplied by the 
  branching ratio of the leptonic channel,
  and nuclear modification of
  \UpsAll mesons, the ground states and the excited states
  separately, in 0--60\% U+U collisions as well as in each
  centrality bin. The uncertainties are listed statistical first and systematic second. The statistical uncertainties from the {\it{p+p}} reference, not included in the table, are 12.7\%, 13.0\% and 30\% for the
  \UpsAll, \UpsOne and \UpsExc respectively.
  There is an additional 11\% common normalization uncertainty on \Raa from
  the {\it{p+p}} luminosity estimation~\cite{Adamczyk:2013poh}.}
\end{table}
Table~\ref{tab:spectra} lists the cross-sections in the given \pT ranges for \UpsAll and \UpsOne{}.
\begin{table}
\center
\begin{tabular}{| c | c | c |}
\hline
states & \pT{} (GeV/$c$) & $\Bee\times\frac{d^2 \sigma^\Upsilon_{AA}}{d\pT{}dy}$ $\left(\frac{\mub}{\mathrm{GeV}/c}\right)$ \\
\hline
\hline
\multirow{3}{*}{\UpsAll}
 & 0--2 & $1.40 \pm 0.49^{+0.36}_{-0.23}$ \\
 & 2--4 & $1.96 \pm 0.51^{+0.42}_{-0.43}$ \\
 & 4--10 & $0.53 \pm 0.77^{+0.20}_{-0.11}$ \\
\hline
\multirow{4}{*}{\UpsOne}
 & 0--2 & $1.30 \pm 0.39^{+0.28}_{-0.22}$ \\
 & 2--4 & $1.61 \pm 0.43^{+0.35}_{-0.35}$ \\
 & 4--10 & $0.30 \pm 0.38^{+0.17}_{-0.05}$ \\
\hline
\end{tabular}
\caption{\label{tab:spectra}Cross-sections multiplied by the branching ratio of the leptonic channel, in given \pT{} ranges for the \UpsAll and \UpsOne states in 0--60\% U+U collisions.}
\end{table}
The \pT{} spectrum is well described by a Boltzmann distribution with a slope parameter of 
$p_0^\UpsAll=(1.37\pm0.20^{+0.03}_{-0.07})$ GeV/$c$ and 
$p_0^\UpsOne=(1.22\pm0.15\pm^{+0.04}_{-0.05})$ GeV/$c$.  
These values are consistent with the interpolation from {\it{p+p}} data within uncertainties.

\begin{figure}
  \includegraphics[trim=3mm 0 12mm 0,clip,width=\linewidth]{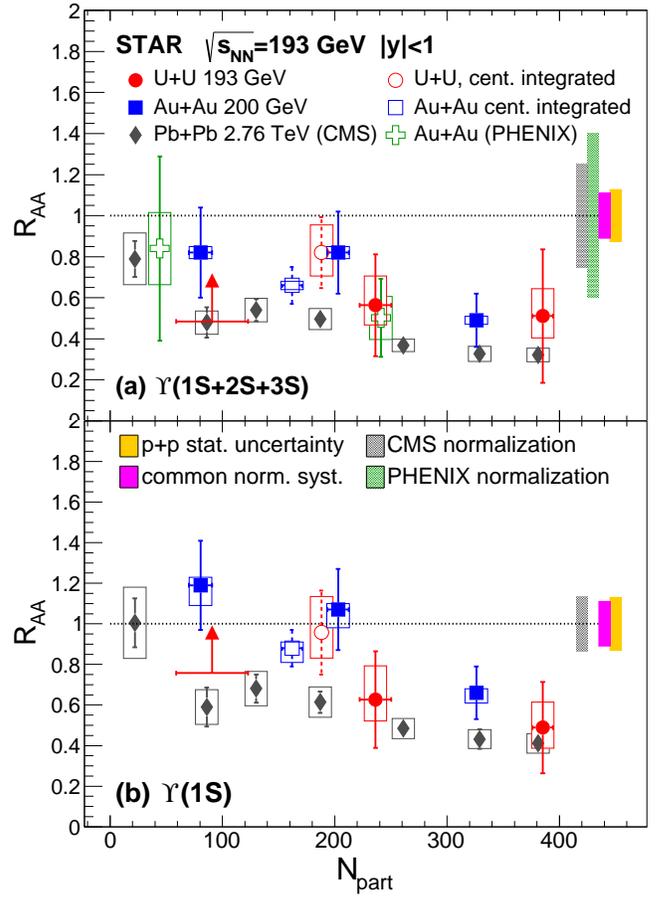}
  \caption{\label{fig:raa-data} {\it (Color online)}
   \UpsAll (a) and \UpsOne (b) \Raa vs. \Npart in $\sqrt{s_{NN}}=193$ GeV U+U collisions
   (solid circles), compared to 200 GeV RHIC Au+Au (solid squares~\cite{Adamczyk:2013poh} and hollow crosses~\cite{Adare:2014hje}), and 2.76 TeV LHC Pb+Pb data
   (solid diamonds~\cite{Chatrchyan:2012lxa}). A 95\% lower confidence bound is indicated for the 30-60\% centrality U+U data (see text). Each point is plotted at the center of its bin. Centrality integrated (0-60\%) U+U and Au+Au data are also shown as open circles and squares, respectively.}
\end{figure}

The \UpsAll and \UpsOne nuclear modification factors as a function of \Npart are shown in Fig.~\ref{fig:raa-data}, and compared to the nuclear modification factor in Au+Au data at $\sqrt{s_{NN}}=200$ GeV from STAR~\cite{Adamczyk:2013poh} at $|\eta|<1$, PHENIX~\cite{Adare:2014hje} at $|\eta|<0.35$, and in Pb+Pb data measured by CMS at $\sqrt{s_{NN}}=2.76$ TeV via the $\Upsilon\rightarrow\mu^+\mu^-$ channel within $|\eta|<2.4$~\cite{Chatrchyan:2012lxa}.
%
%
%
\begin{figure}
  \includegraphics[trim=3mm 0 12mm 0,width=\linewidth]{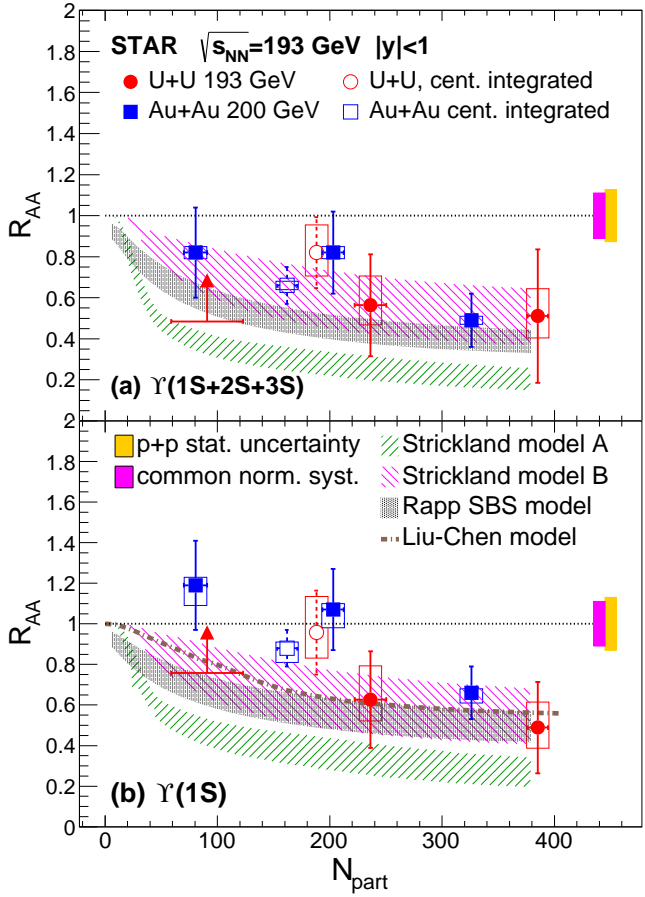}
  \caption{\label{fig:raa-model} {\it (Color online)}
   \UpsAll (a) and \UpsOne (b) \Raa vs. \Npart in $\sqrt{s_{NN}}=193$ GeV U+U collisions (solid circles), compared to different models~\cite{Emerick:2011xu,Strickland:2011aa,Liu:2010ej}, described in the text.
   The 95\% lower confidence bound is indicated for the 30-60\% centrality U+U data (see text). Each point is plotted at the center of its bin. Centrality integrated (0-60\%) U+U and Au+Au data are also shown as open circles and squares, respectively.}
\end{figure}
The data points in the 30-60\% centrality bin have large statistical and systematical uncertainties, providing little constraint on \Raa. In Figs.~\ref{fig:raa-data} and \ref{fig:raa-model} we therefore only show the 95\% lower confidence bound for these points, derived by quadratically adding statistical and point-to-point systematic uncertainties.
The \Raa values measured in all \Npart bins for the \UpsAll, \UpsOne and \UpsExc states are summarized in Table~\ref{tab:result}.
Note that the \UpsOne results are not corrected for feed-down from the excited states.
%
\begin{figure}
  \includegraphics[trim=0 0 17mm 11mm,width=\linewidth]{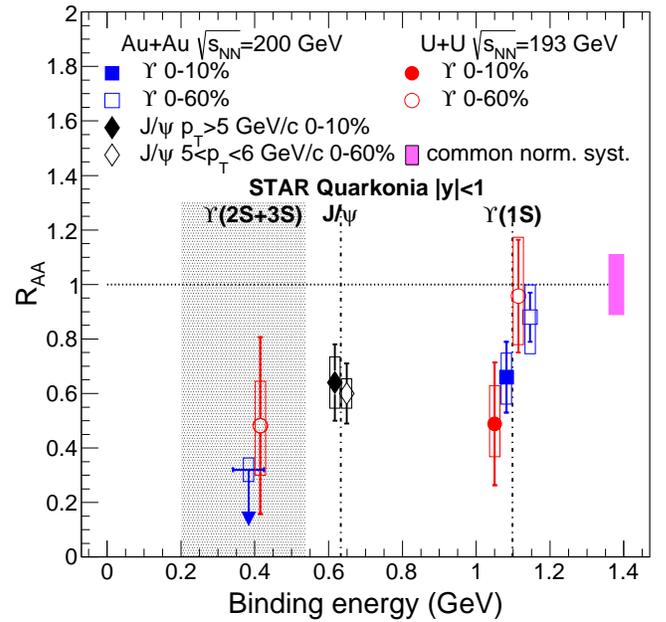}
  \caption{\label{fig:binding} {\it (Color online)}
   Quarkonium \Raa versus binding energy in Au+Au and U+U
   collisions. Open symbols represent 0-60\% centrality data, filled symbols are for 0-10\% centrality.
   The \Ups measurements in U+U collisions are denoted by red points. In the case of Au+Au collisions, the \UpsOne measurement is denoted by a blue square, while for the \UpsExc states, a blue horizontal line indicates a 95\% upper confidence bound. The black diamonds mark the high-\pT{} \Jpsi measurement. The vertical lines represent nominal binding energies for the \UpsOne and \Jpsi, calculated based on the mass defect, as $2m_D-m_{\Jpsi}$ and $2m_B-m_\Ups$, respectively (where $m_X$ is the mass of the given meson $X$)~\cite{Satz:2006kba}. The shaded area spans between the binding energies of \UpsTwo and \UpsThree. The data points are slightly shifted to the left and right from the nominal binding energy values to improve their visibility.
 } 
\end{figure}

The trend marked by the Au+Au $\Raa(\Npart)$ points is augmented by the U+U data.
We observe neither a significant difference between the results in any of the centrality classes, nor do we find any evidence of a sudden increase in suppression in central U+U  compared to the central Au+Au data, although the precision of the current measurement does not exclude a moderate drop in \Raa.
Assuming that the difference in suppression between the Au+Au and U+U collisions
is small, the two data sets can be combined. We carry out the unification using the BLUE method~\cite{Valassi:2003mu,Nisius:2014wua} with the conservative assumption that all common systematic uncertainties are fully correlated. We find that \UpsOne{} production is significantly suppressed in central heavy-ion collisions at top RHIC energies, but this suppression is not complete: 
$\Raa^\UpsOne=0.63 \pm 0.16 \pm 0.09$ where the first uncertainty includes both the unified statistical and systematic errors and the second one is the global scaling  uncertainty from the {\it p+p} reference.
While both the RHIC and LHC data show suppression in the most central bins, $\Raa^\UpsOne$ is slightly, although not significantly, higher in RHIC semi-central collisions than in the LHC.
In the Au+Au data, the \UpsExc excited states have been found to be strongly suppressed, and an upper limit $\Raa^\UpsExc < 0.32$ was established. The \UpsExc suppression observed in U+U data is consistent with this upper limit.

In Fig.~\ref{fig:raa-model} we compare STAR measurements to different
theoretical models~\cite{Emerick:2011xu,Strickland:2011aa,Liu:2010ej}.
An important source of uncertainty in model calculations for
quarkonium dissociation stems from the unknown nature of the in-medium potential between the quark-antiquark pairs. Two limiting cases that are
often used are the internal-energy-based heavy quark potential
corresponding to a strongly bound scenario (SBS), and the free-energy-based
potential corresponding to a more weakly bound scenario (WBS)~\cite{Grandchamp:2005yw}.
The model of Emerick, Zhao and Rapp~\cite{Emerick:2011xu} includes
CNM effects, dissociation of bottomonia in the hot
medium (assuming a temperature $T=330$ MeV) and
regeneration for both the SBS and WBS scenarios. The Strickland-Bazow model~\cite{Strickland:2011aa} calculates dissociation in the medium in both a free-energy-based ``model A'' and an internal-energy-based ``model B'', with an initial central temperature
$428<T<442$ MeV. The model of Liu {\it et al.}~\cite{Liu:2010ej} uses
an internal-energy-based potential and an input temperature $T=340$
MeV.
In Fig.~\ref{fig:raa-model} we show all three internal-energy-based models together with the ``model A'' of Ref.~\cite{Strickland:2011aa} as an example for the free-energy-based models. The internal-energy-based models generally describe RHIC data well within the current uncertainties, while the free-energy-based models tend to underpredict the \Raa especially for the \UpsOne.

Figure~\ref{fig:binding} shows the \Raa versus binding energy of
\UpsOne and \UpsExc states~\cite{Satz:2006kba} in U+U and Au+Au collisions. 
The results are also compared to high-\pT
$J/\psi$ in Au+Au collisions~\cite{Adamczyk:2012ey}. This comparison is motivated by the expectation from model calculations, e.g.\ that in Ref.~\cite{Liu:2009nb}, that charm recombination is moderate at higher momenta. 
Recent measurements at the LHC~\cite{Khachatryan:2016xxp,Khachatryan:2016ypw} indicate that the suppression of the \Ups production, as well as that of the prompt \Jpsi in the $\pT>5$ GeV/$c$ range, is rather independent of the momentum of the particle.
Contrary to earlier assumptions~\cite{Xu:1995eb,Liu:2012zw}, no noticeable \pT or rapidity dependence was observed. 
However, the non-prompt \Jpsi production~\cite{Khachatryan:2016ypw},
originating dominantly from $B$ meson decays, does show
a clear \pT dependence ~\cite{Adamczyk:2012ey}.
This affects the pt-dependence of inclusive \Jpsi production, especially at high-\pT.
Our current data does not have sufficient statistics to study the \pT
dependence of the \Ups in detail and to verify whether the observations
at the LHC also hold at RHIC energies.
The results in U+U collisions are consistent with the Au+Au measurements as well as with the expectations from the sequential melting hypothesis.

\section{Summary}

We presented mid-rapidity measurements of inclusive bottomonium production in U+U collisions at $\sqrt{s_{NN}}=193$ GeV. The cross-section is $\Bee \times(d\sigma^\Ups_{AA}/dy)=
4.27 \pm 0.90^{+0.90}_{-0.82}
$ \mub{} for the \UpsAll, and $\Bee \times(d\sigma^\UpsOne_{AA}/dy)=
3.55 \pm 0.77^{+0.80}_{-0.66}
$ \mub{} for the separated \UpsOne state.

The present measurements increased the range of the number of participants in the collision compared to the previous Au+Au measurements by approximately 20\%. A significant suppression is observed in central U+U data for both the \UpsAll{} ($\Raa^\Ups=0.51 \pm 0.32^{+0.13}_{-0.11} \pm0.08$, where the first uncertainty reflects the statistical error, the second the overal systematic uncertainty, and the third the uncertainty from the {\it p+p} reference) and 
\UpsOne{} ($\Raa^\UpsOne=0.49 \pm 0.23^{+0.12}_{-0.10} \pm0.09$), which
consolidates and extends the previously observed
\Raa{}(\Npart) trend in Au+Au collisions.
The data from 0-60\% central U+U
collisions is consistent with a strong suppression of the 
\UpsExc states, which has also been observed in Au+Au collisions.
Comparison of the suppression patterns {from Au+Au and U+U data to different models} favors an internal-energy-based quark potential scenario.

\section*{Acknowledgement}

We thank the RHIC Operations Group and RCF at BNL, the NERSC Center at LBNL, 
and the Open Science Grid consortium for providing resources and support. This work was supported in part by the Office of Nuclear Physics within the U.S. DOE Office of Science, the U.S. NSF, the Ministry of Education and Science of the Russian Federation, NSFC, CAS, MoST and MoE of China, the National Research Foundation of Korea, NCKU (Taiwan), GA and MSMT of the Czech Republic, FIAS of Germany, DAE, DST, and UGC of India, the National Science Centre of Poland, National Research Foundation, the Ministry of Science, Education and Sports of the Republic of Croatia, and RosAtom of Russia.

\bibliographystyle{elsarticle-num}

\end{document}